\documentclass[conference]{IEEEtran}
\IEEEoverridecommandlockouts
\usepackage{cite}
\usepackage{amsmath,amssymb,amsfonts}
\usepackage{algorithmic}
\usepackage{graphicx}
\usepackage{textcomp}
\usepackage{xcolor}
\usepackage{todonotes}
\def\BibTeX{{\rm B\kern-.05em{\sc i\kern-.025em b}\kern-.08em
    T\kern-.1667em\lower.7ex\hbox{E}\kern-.125emX}}
\begin{document}
\newcommand\copyrighttext{%
  \footnotesize \textcopyright 2020 IEEE. Personal use of this material is permitted. Permission from IEEE must be obtained for all other uses, in any current or future media, including reprinting/republishing this material for advertising or promotional purposes,creating new collective works, for resale or redistribution to servers or lists, or reuse of any copyrighted component of this work in other works.
  Conference: 2020 Ural Symposium on Biomedical Engineering, Radioelectronics and Information Technology (USBEREIT). Preprint available here: https://arxiv.org/abs/1911.09731}
\newcommand\copyrightnotice{%
\begin{tikzpicture}[remember picture,overlay]
\node[anchor=south,yshift=10pt] at (current page.south) {\fbox{\parbox{\dimexpr\textwidth-\fboxsep-\fboxrule\relax}{\copyrighttext}}};
\end{tikzpicture}%
}

\newpage

\title{Phase Mapping for Cardiac Unipolar Electrograms with Neural Network Instead of Phase Transformation\\
\thanks{The reported study was funded by RFBR, according to the research project No. 18-31-00401. Development of the mathematical models is supported by IIF UrB RAS theme \#AAAA-A19-119070190064-4, RF Government Act \#211 of March 16, 2013, the Program of the Presidium RAS.}
}

\author{\IEEEauthorblockN{1\textsuperscript{st} Konstantin Ushenin, Vladimir Sholokhov, Tatyana Nesterova, Dmitry Shmarko}
\IEEEauthorblockA{\textit{Institute of Natural Sciences and Mathematics} \\
\textit{Ural Federal University}\\
konstantin.ushenin@urfu.ru}
}

\author{\IEEEauthorblockN{Konstantin Ushenin}
\IEEEauthorblockA{\textit{Institute of Natural Sciences} \\
\textit{Ural Federal University}\\
Ekaterinburg, Russia \\
konstantin.ushenin@urfu.ru}
\and
\IEEEauthorblockN{Tatyana Nesterova}
\IEEEauthorblockA{\textit{Laboratory of Mathematical Physiology} \\
\textit{Institute of Immunology and Physiology}\\
Ekaterinburg, Russia \\
tatiannesterova@gmail.com}
\and
\IEEEauthorblockN{Dmitry Smarko}
\IEEEauthorblockA{\textit{Laboratory of Mathematical Physiology} \\
\textit{Institute of Immunology and Physiology}\\
Ekaterinburg, Russia \\
d.shmarko@yandex.ru}
\and
\IEEEauthorblockN{Vladimir Sholokhov}
\IEEEauthorblockA{\textit{Institute of Natural Sciences} \\
\textit{Ural Federal University}\\
Ekaterinburg, Russia \\
vdsholokhov@yandex.ru}
}

\maketitle
\copyrightnotice
\begin{abstract}
A phase mapping is an approach to processing signals of electrograms that are recorded from the surface of cardiac tissue. The main concept of the phase mapping is an application of the phase transformation with the aim to obtain signals with useful properties. In our study, we propose to use a simple sawtooth signal instead of a phase of a signal for processing of electrogram data and building of the phase maps. We denote transformation that can provide this signal as a phase-like transformation (PLT). PLT defined via a convolutional neural network that is trained on a dataset from computer models of cardiac tissue electrophysiology. The proposed approaches were validated on data from the detailed personalized model of the human torso electrophysiology. This paper includes visualization of the phase map based on PLT and shows the applicability of the proposed approaches in the analysis of the complex non-stationary periodic activity of the excitable cardiac tissue.
\end{abstract}


\section{Introduction}

A unipolar electrogram is a popular method for invasive electrophysiological studies in cardio-surgery  \cite{tedrow2011recording}. Cardiac mapping is a modern extension of a unipolar electrogram analysis that presents cardiac electrophysiology in the format of maps or video maps.

The most complex processing is required for the presentation of the periodic and non-stationary periodic activity of myocardium (cardiac muscle tissue). This activity usually observed during the ventricular tachycardia or atrial flutter. A phase mapping is the most common approach for processing of such type of data. These approaches were widely used in biological \textit{in vitro} experiments \cite{bray2002considerations,nash2006evidence,umapathy2010phase}, and lately was translated to clinical practice \cite{kuklik2015reconstruction, vijayakumar2016methodology,dubois2016electrocardiographic}.

\begin{figure}[t]
\centering
\includegraphics[width=0.47\textwidth]{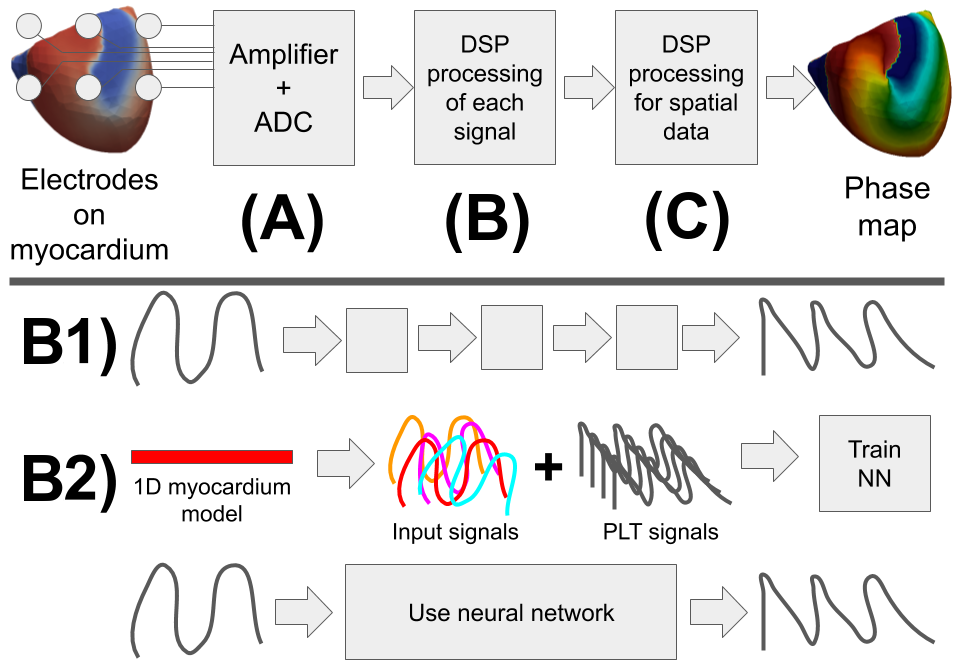}
\caption{The main idea of the current paper. A phase mapping is an approach to process data from unipolar electrodes placed on the myocardium. It is shown on the top row (A, B, C). Some parts of the phase mapping require the processing of signals from each electrode (B). The signal processing pipeline here is usually a sequence of mathematical transformations (B1). We propose to use a computer model of a 1D myocardial strand for the generation of training datasets for the neural networks (B2). The trained neural network can replace all steps of the pipeline. Abbreviations: analog-to-digital converter (ADC), digital signal processing (DSP), Neural network (NN).}
\label{fig:idea2}
\end{figure}

The phase mapping approach includes two parts. The first part is the phase transformation of the signals that are recorded from several leads, and the second part is an interpolation and spatial analysis of all phase signals (see Fig. \ref{fig:idea2}). The phase transformation usually based on a shift in time or Hilbert transform \cite{dubois2016electrocardiographic}. Also, several alternative approaches were proposed \cite{kuklik2015reconstruction}.

\begin{figure}[t]
\centering
\includegraphics[width=0.40\textwidth]{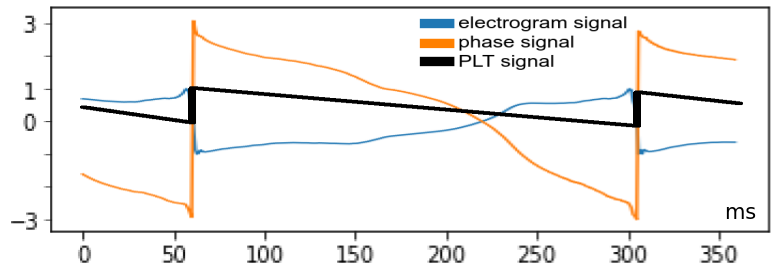}
\caption{Explanation of phase-like transformation (PLT). In phase map processing, the original electrogram signal (blue line; visualized after normalization) is transformed to the phase signal (orange line). We propose to use the sawtooth signal (black line) that has basic properties of phase signal but is significantly simpler.}
\label{fig:idea}
\end{figure}

Here, we are aiming to replace the phase transformation on the more robust approach. We propose to use a sawtooth signal with [0,1] range of values. These signals should have breaks that related to depolarization of transmembrane potential in cardiomyocytes in point of measurements. Between two depolarizations, the signal should linearly decrease from 1 to 0 value. Fig. \ref{fig:idea} shows the proposed signal and its relationship with the cardiomyocyte transmembrane potential.

This signal should be obtained from unipolar electrogram with some transformation. We named this transformation as the phase-like transformation (PLT). In contradiction to previously proposed approaches \cite{bray2002considerations,nash2006evidence,umapathy2010phase, kuklik2015reconstruction, vijayakumar2016methodology,dubois2016electrocardiographic} we do not define this transformation as series of mathematical operations (see Fig. \ref{fig:idea2}  (B1)). We define the PLT via training of neural network on series of simple 1D model of myocardial electrophysiology (see Fig. \ref{fig:idea2} (B2)). 

For proof-of-concept, we test the proposed idea with the more complex and detailed model of the human heart electrophysiology.
\section{Methods}

In this study, a simple 1D model of the myocardium provides a series of action potentials. These action potentials converted to unipolar signals with a simple equation. Obtained signals separated into the training and validation dataset. Then, the neural network is trained on these datasets. The complex and realistic model of the human torso electrophysiology provides action potentials and extracellular potential, and the last one is very close to the real unipolar signals. Extracellular potentials are processed with the neural network for build phase maps based on the PLT transformation. In the end, these maps are compared with true electrophysiological activity in the myocardium that is shown via a map of the action potentials from the realistic model in a fixed point in time.

\textit{Idealized 1D models of myocardial tissue} were computed with the monodomain equation. The 1D strand contained 1024 points, and the activation point was located in 128 nodes from one side of the strands. TNNP06 \cite{ten2006alternans} described electrophysiological activity of ventricular cardiomyocytes respectively.

\textit{Unipolar electrograms from 1D strand} were obtained with the following formula \cite{weinberg2008representation}:

\begin{equation}
    \phi(x') = - \kappa \int_x \frac{\partial V}{\partial x}\cdot \frac{\partial}{\partial x}\big(\frac{1}{\sqrt{(x-x')^2+h^2}}\big) dx,
\end{equation}
where $V$ is the transmembrane potential in point $x$, $\phi$ is an extracellular potential or signal from a unipolar catheter, $h$ is the height of the catheter above the 1D strand, and $x'$ is an electrode position.

The coefficient $\kappa$ and a voltage in absolute physical values are not important for our study because we normalized results using the division of each signal to their maximal absolute amplitude.

\begin{figure*}[t]
\centering
\includegraphics[width=0.85\textwidth]{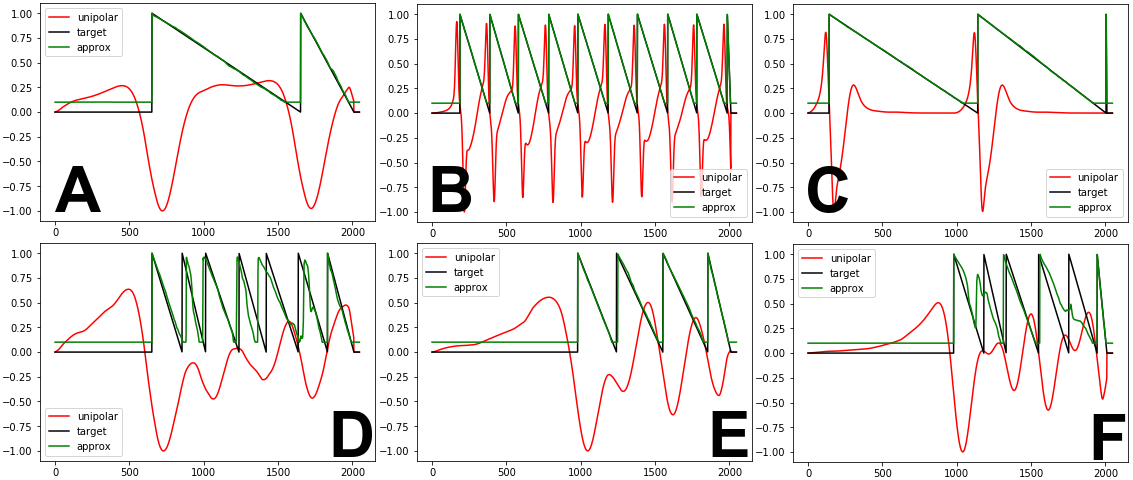}
\caption{Examples of electrogram signals, target PLT signals, and PLT signals provided in the output of the neural network. The top row (A, B, C) shows a signal for periodic activation of the strand, and the bottom row (D, E, F) shows signals for strand activation on high frequency, which causes action potential alternations. Cases A, B, C, D, E were correctly processed by the neural network, and case F includes artifacts.}
\label{fig:examples}
\end{figure*}

Training and validation datasets were generated with a variation of the following 1D model parameters: the stimulation frequency ($FR=\{2000, 1000, 500, 300, 200\}$ Hz), conduction velocity ($CV=\{10,20,40,80\}$).  Parameters of the cardiomyocytes in the 1D strand were taken from the original article \cite{ten2006alternans} without changes. Also, the following parameters of the (1) equation were variated: height of electrode over the strand ($h=\{5,10,20,50,80\}$) and position of electrode along the strand ($x'=\{448, 512, 640\}$). As shown in Fig. \ref{fig:examples}, used ranges of parameter variation provide significantly different signals for analysis and cover all possible signal shapes, available from the real recordings.

PLT signals for training were generated from action potential using a 0 mV threshold level as criteria for the depolarization phase (break of the function). Examples of PTL signals are shown in Fig. \ref{fig:examples}. Thus, the full datasets of simple model results contained 300 signals with 4096 ms lengths and 1000 Hz frequency of discretization.  The training and validation datasets respectively contain 150 (50\%) and 150 (50\%) signals.

\textit{The test dataset} was generated by a detailed personalized finite element model of two ventricles and the torso. Model geometry was based on computed tomography data of one patient. The torso includes regions of the heart, lungs, blood in heart chambers, and spinal cord. Each torso region had realistic conductivity, according to \cite{keller2010ranking}. The heart included realistic conduction anisotropy that was introduced with a rule-based approach and realistic heterogeneity of current transmembrane densities \cite{keller2012influence}. The TNNP06 model \cite{ten2006alternans} performed a realistic simulation of cardiomyocytes electrophysiology. A bidomain model with bath described excitation wave propagation and the torso electrophysiology.  We initiate a spiral wave using the S1S2 protocol to provide the realistic extracellular potential for ventricular arrhythmia of the reentry type.  Each point of the heart surface mesh provides one signal for the test dataset. Thus, the entire model provides 34354 signals with 4096 ms length.

The described approach is one of the most realistic ways for the simulation of electrophysiology in both ventricles and the torso. In particular, this approach correctly includes the far-field effect. The used model was verified against clinical data of electrocardiography with 224 leads during activation of the myocardium from a point \cite{ushenin67role}. We suppose that model complexity and a wide representation of physiological features make the model suitable for the generation of the test dataset. This dataset was used only with the neural network that is trained to process the signals from the ventricular myocardium.

\begin{figure}[h]
\centering
\includegraphics[width=0.30\textwidth]{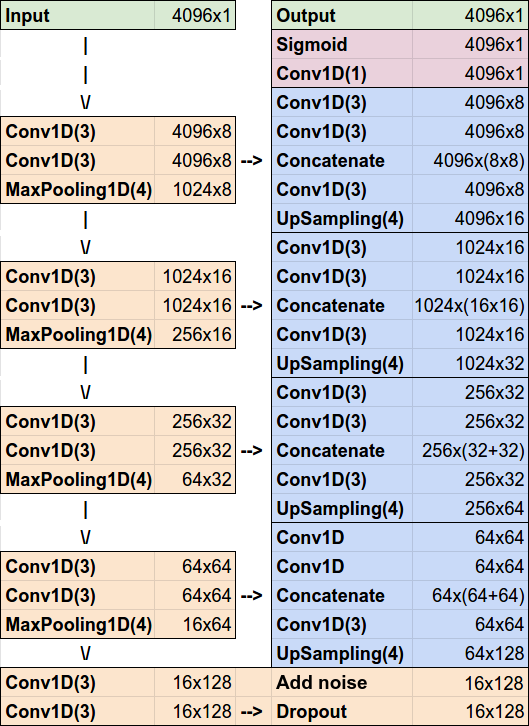}
\caption{Architecture of the neural network.}
\label{fig:nn}
\end{figure}

\textit{Convolutional neural network for processing} was adapted from the U-Net architecture for biomedical image segmentation \cite{ronneberger2015u}. Fig. \ref{fig:nn} presents our modifications. The size of all convolution kernels was replaced from 3x3 to 3x1 elements. The size of the pooling and up-sampling layers was replaced from 2x2 to 4x1 with an aim to increase the perception field of NN. The number of neurons in all layers was proportionally increased for the processing of input vectors with 4096 elements. Also, we add Dropout (30\%) and Gaussian noise layer (mean=0, std.=0.2) after the U-Net narrow layer for improving NN robustness. The loss function was a sum of the mean absolute error and mean squared error with equal weights. We used the ADAM method of optimization with learning rate reduction on the plateau of validation loss. The NN process unipolar signals from windows with 4096 values and provides a PLT signal as an output with 4096 values.

\section{Results}

\textit{The model of cardiac tissue} provides a set of excitation waves for a normal healthy myocardium (Fig. \ref{fig:examples} (A-B)) and additionally provide cases with alternance of  action potentials (Fig. \ref{fig:examples} (D-F)). The last ones appeared under high stimulation frequency at the 1D strand.

Idealized models provide a set of unipolar electrograms with significant differences between cases. The majority of signals were similar to clinically observed electrophysiological behaviour for sino-atrial rhythm (Fig. \ref{fig:examples} (A, C), \cite{tedrow2011recording}), flutter or fibrillation (Fig. \ref{fig:examples} (B, D--F), \cite{konings1997configuration}). However, some signals look atypical, and we suppose that the dataset covers both real and unobtainable cases of electrogram signals.

\begin{figure*}[t]
\centering
\includegraphics[width=0.85\textwidth]{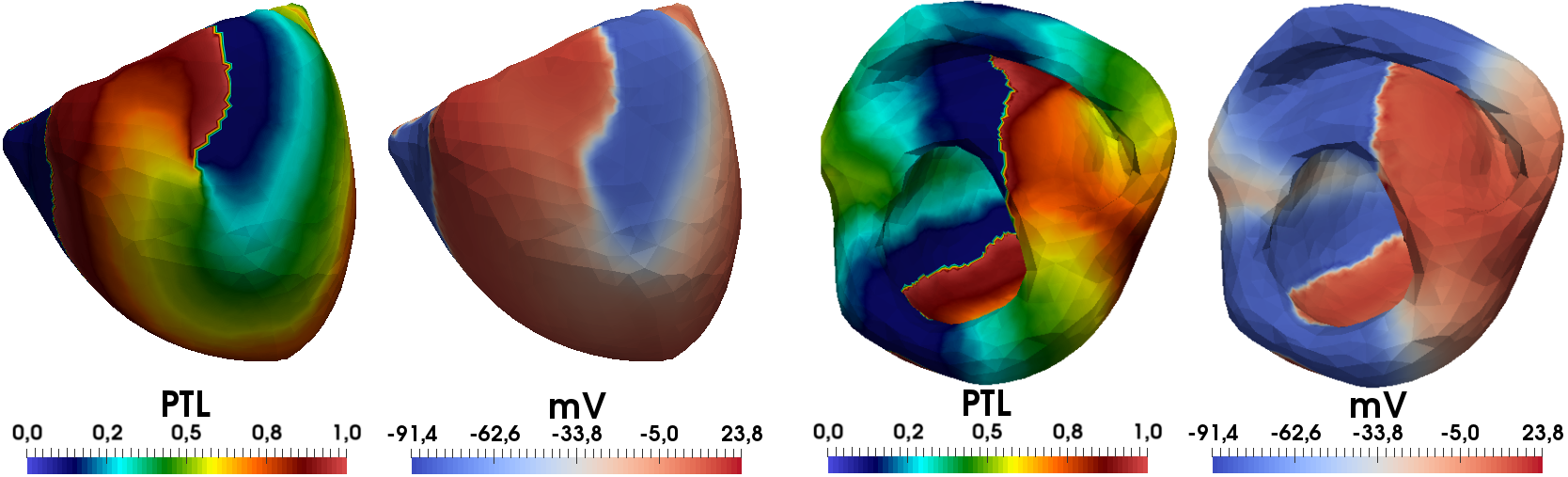}
\caption{Phase maps based on PLT signals and action potential in a fixed point in time. Data were obtained in the detailed computer model of the ventricular tachycardia.}
\label{fig:maps}
\end{figure*}


\begin{table}[h]
\centering
\caption{Metrics of neural network performance at 150 epoch. Mean absolute error (MAE), mean square error (MSE).}
\label{tbl:losses}
\begin{tabular}{lc|cc}
            & Num. of cases &  \multicolumn{2}{c}{Ventricles}   \\
            &  & MAE & MSE   \\ 
Train (1D) & 150  &  0.0250 & 0.0021 \\
Validation (1D) & 150  &  0.0327 & 0.0024 \\
Test (3D) & 34354 &  0.0888 & 0.0266 \\
\end{tabular}
\end{table}
NN training requires 100+ epochs for reaching of a loss plateau. Table \ref{tbl:losses} shows the training process and the final value of the loss, respectively. We manually analyzed shapes of the sawtooth signals and counted a number of wrongly detected action potential upstrokes in the validation dataset. Only 10 out of 859 (1.16\%) upstrokes were incorrectly identified.

Qualitative analysis of NN outputs is presented in Fig. \ref{fig:examples}. NN perfectly processes any periodic signals and the major part of non-stationary periodic signals (see Fig. \ref{fig:examples} (A--E)). However, the model makes an error with some non-stationary periodic signals that contain double peaks with close placement (see Fig. \ref{fig:examples} (F)). The model cannot reach zero levels in zones without electrophysiological activity and in the ends of the sawtooth shapes.

The result of the NN application on the realistic personalized model of human electrophysiology is presented in Fig. \ref{fig:maps}. As can be seen, the obtained quality of the phase map clearly reveals the rotor core and the front of the excitation waves. However, the value of the mean absolute error metric here is at three and half times higher than the ones for 1D strand models (see Table \ref{tbl:losses}).

\section{Discussion and Conclusion}

In our study, we propose phase-like transformation (PLT) for processing unipolar electrograms and the method of its definition via the convolutional neural network that is trained on a set of generated data from the numerical experiments.

The proposed transformation provides signals with desirable properties as we planned in the beginning. It can reveal complex non-stationary periodic behavior in the myocardium and is applicable for the building of phase maps (see Fig. \ref{fig:maps}). 

Our approach to transformation definition significantly differs from the method that was proposed before. It does not require a manual choice of signal transformations and filters with good basic properties. Instead of that, they require proper choice and tuning of several models of the physiological process. Loss functions here are a direct way for assessment of the algorithm able to process complex data.

We suppose that the proposed approach has a wide area of application. It may be applied to the processing of unipolar electrograms from the unipolar catheter, multi-leads catheter, and balloons, microelectrode arrays, invasive and non-invasive systems of cardiac mapping \cite{kalinin2019solving,revishvili2015validation}.

\bibliographystyle{IEEEtran}
\bibliography{bibliography}

\begin{thebibliography}{10}
\providecommand{\url}[1]{#1}
\csname url@samestyle\endcsname
\providecommand{\newblock}{\relax}
\providecommand{\bibinfo}[2]{#2}
\providecommand{\BIBentrySTDinterwordspacing}{\spaceskip=0pt\relax}
\providecommand{\BIBentryALTinterwordstretchfactor}{4}
\providecommand{\BIBentryALTinterwordspacing}{\spaceskip=\fontdimen2\font plus
\BIBentryALTinterwordstretchfactor\fontdimen3\font minus
  \fontdimen4\font\relax}
\providecommand{\BIBforeignlanguage}[2]{{%
\expandafter\ifx\csname l@#1\endcsname\relax
\typeout{** WARNING: IEEEtran.bst: No hyphenation pattern has been}%
\typeout{** loaded for the language `#1'. Using the pattern for}%
\typeout{** the default language instead.}%
\else
\language=\csname l@#1\endcsname
\fi
#2}}
\providecommand{\BIBdecl}{\relax}
\BIBdecl

\bibitem{tedrow2011recording}
U.~B. Tedrow and W.~G. Stevenson, ``Recording and interpreting unipolar
  electrograms to guide catheter ablation,'' \emph{Heart Rhythm}, vol.~8,
  no.~5, pp. 791--796, 2011.

\bibitem{bray2002considerations}
M.-A. Bray and J.~P. Wikswo, ``Considerations in phase plane analysis for
  nonstationary reentrant cardiac behavior,'' \emph{Physical Review E},
  vol.~65, no.~5, p. 051902, 2002.

\bibitem{nash2006evidence}
M.~P. Nash, A.~Mourad, R.~H. Clayton, P.~M. Sutton, C.~P. Bradley, M.~Hayward,
  D.~J. Paterson, and P.~Taggart, ``Evidence for multiple mechanisms in human
  ventricular fibrillation,'' \emph{Circulation}, vol. 114, no.~6, pp.
  536--542, 2006.

\bibitem{umapathy2010phase}
K.~Umapathy, K.~Nair, S.~Masse, S.~Krishnan, J.~Rogers, M.~P. Nash, and
  K.~Nanthakumar, ``Phase mapping of cardiac fibrillation,'' \emph{Circulation:
  Arrhythmia and Electrophysiology}, vol.~3, no.~1, pp. 105--114, 2010.

\bibitem{kuklik2015reconstruction}
P.~Kuklik, S.~Zeemering, B.~Maesen, J.~Maessen, H.~J. Crijns, S.~Verheule,
  A.~N. Ganesan, and U.~Schotten, ``Reconstruction of instantaneous phase of
  unipolar atrial contact electrogram using a concept of sinusoidal
  recomposition and hilbert transform,'' \emph{IEEE transactions on biomedical
  engineering}, vol.~62, no.~1, pp. 296--302, 2015.

\bibitem{vijayakumar2016methodology}
R.~Vijayakumar, S.~K. Vasireddi, P.~S. Cuculich, M.~N. Faddis, and Y.~Rudy,
  ``Methodology considerations in phase mapping of human cardiac arrhythmias,''
  \emph{Circulation: Arrhythmia and Electrophysiology}, vol.~9, no.~11, p.
  e004409, 2016.

\bibitem{dubois2016electrocardiographic}
R.~Dubois, A.~Pashaei, J.~Duchateau, and E.~Vigmond, ``Electrocardiographic
  imaging and phase mapping approach for atrial fibrillation: A simulation
  study,'' in \emph{Computing in Cardiology Conference (CinC), 2016}.\hskip 1em
  plus 0.5em minus 0.4em\relax IEEE, 2016, pp. 117--120.

\bibitem{ten2006alternans}
K.~H. Ten~Tusscher and A.~V. Panfilov, ``Alternans and spiral breakup in a
  human ventricular tissue model,'' \emph{American Journal of Physiology-Heart
  and Circulatory Physiology}, vol. 291, no.~3, pp. H1088--H1100, 2006.

\bibitem{weinberg2008representation}
S.~Weinberg, S.~Iravanian, and L.~Tung, ``Representation of collective
  electrical behavior of cardiac cell sheets,'' \emph{Biophysical journal},
  vol.~95, no.~3, pp. 1138--1150, 2008.

\bibitem{keller2010ranking}
D.~U. Keller, F.~M. Weber, G.~Seemann, and O.~Dossel, ``Ranking the influence
  of tissue conductivities on forward-calculated ecgs,'' \emph{IEEE
  Transactions on Biomedical Engineering}, vol.~57, no.~7, pp. 1568--1576,
  2010.

\bibitem{keller2012influence}
D.~U. Keller, D.~L. Weiss, O.~Dossel, and G.~Seemann, ``Influence of $i_{Ks}$
  heterogeneities on the genesis of the t-wave: A computational evaluation,''
  \emph{IEEE Transactions on Biomedical Engineering}, vol.~59, no.~2, pp.
  311--322, 2012.

\bibitem{ushenin67role}
K.~S. Ushenin, A.~Dokuchaev, S.~M. Magomedova, O.~V. Sopov, V.~V. Kalinin,
  O.~Solovyova, and E.~S. SA, ``Role of myocardial properties and pacing lead
  location on ecg in personalized paced heart models,'' \emph{Age}, vol.~67,
  p.~56, 2018.

\bibitem{ronneberger2015u}
O.~Ronneberger, P.~Fischer, and T.~Brox, ``U-net: Convolutional networks for
  biomedical image segmentation,'' in \emph{International Conference on Medical
  image computing and computer-assisted intervention}.\hskip 1em plus 0.5em
  minus 0.4em\relax Springer, 2015, pp. 234--241.

\bibitem{konings1997configuration}
K.~T. Konings, J.~L. Smeets, O.~C. Penn, H.~J. Wellens, and M.~A. Allessie,
  ``Configuration of unipolar atrial electrograms during electrically induced
  atrial fibrillation in humans,'' \emph{Circulation}, vol.~95, no.~5, pp.
  1231--1241, 1997.

\bibitem{kalinin2019solving}
A.~Kalinin, D.~Potyagaylo, and V.~Kalinin, ``Solving the inverse problem of
  electrocardiography on the endocardium using a single layer source,''
  \emph{Frontiers in physiology}, vol.~10, p.~58, 2019.

\bibitem{revishvili2015validation}
A.~S. Revishvili, E.~Wissner, D.~S. Lebedev, C.~Lemes, S.~Deiss, A.~Metzner,
  V.~V. Kalinin, O.~V. Sopov, E.~Z. Labartkava, A.~V. Kalinin \emph{et~al.},
  ``Validation of the mapping accuracy of a novel non-invasive epicardial and
  endocardial electrophysiology system,'' \emph{Europace}, p. euu339, 2015.

\end{thebibliography}

\end{document}